\documentclass[]{revtex4}

\usepackage{gensymb}
\usepackage{amsmath}
\usepackage{amssymb}
\usepackage{bm} 
\usepackage{color}
\usepackage[linkcolor=blue, colorlinks=true, urlcolor=blue,
citecolor=red]{hyperref}
\usepackage{graphicx}
\usepackage[caption=false]{subfig}
\usepackage{bbold}
\usepackage{verbatim}
\usepackage{appendix}
\usepackage{amsthm}
\theoremstyle{plain}

\usepackage[normalem]{ulem}

\usepackage{lettrine}

\newcommand{\ket}[1]{\left| #1 \right>} 
\newcommand{\bra}[1]{\left< #1 \right|} 
\newcommand{\bras}[2]{{}_{#2}\hspace*{-0.2mm}\langle{#1}|}
\newcommand{\kets}[2]{|{#1}\rangle_{#2}\hspace*{-0.2mm}}
\newcommand{\braket}[2]{\left< #1 \vphantom{#2} \right|
	\left. #2 \vphantom{#1} \right>} 
\newcommand{\ketbra}[2]{\ket{#1}\hspace*{-0.mm}\bra{#2}}
\newcommand{\ketbras}[3]{\ket{#1}_{#3}\hspace*{-0.mm}\bra{#2}}

\newcommand{\up}{\ket{\uparrow}}
\newcommand{\upp}{\ket{\uparrow\uparrow}}
\newcommand{\down}{\ket{\downarrow}}
\newcommand{\downn}{\ket{\downarrow\downarrow}}
\newcommand{\updown}{\ket{\uparrow\downarrow}}
\newcommand{\downup}{\ket{\downarrow\uparrow}}
\newcommand{\bup}{\bra{\uparrow}}
\newcommand{\bdown}{\bra{\downarrow}}
\newcommand{\Id}{\mathbb{1}}
\newcommand{\Tr}{\mathrm{\text{Tr}}}
\newcommand{\dep}[1]{\left[#1\right]}

\begin{document}

\title{Two-walker discrete-time quantum walks on the line with percolation}

\author{L. Rigovacca}
\affiliation{Quantum Optics and Laser Science Group, Imperial College London, Blackett Laboratory, SW7 2AZ London, UK}   
 
\author{C. Di Franco}
\affiliation{Quantum Optics and Laser Science Group, Imperial College London, Blackett Laboratory, SW7 2AZ London, UK}

\begin{abstract}
{\bf One goal in the quantum-walk research is the exploitation of the intrinsic quantum nature of multiple walkers, in order to achieve the full computational power of the model. Here we study the behaviour of two non-interacting particles performing a quantum walk on the line when the possibility of lattice imperfections, in the form of missing links, is considered.
We investigate two regimes, statical and dynamical percolation, that correspond to different time scales for the imperfections evolution with respect to the quantum-walk one.  
By studying the qualitative behaviour of three two-particle quantities for different probabilities of having missing bonds, we argue that the chosen symmetry under particle-exchange of the input state strongly affects the output of the walk, even in noisy and highly non-ideal regimes.
We provide evidence against the possibility of gathering information about the walkers indistinguishability from the observation of bunching phenomena in the output distribution, in all those situations that require a comparison between averaged quantities.
Although the spread of the walk is not substantially changed by the addition of a second particle, we show that the presence of multiple walkers can be beneficial for a procedure to estimate the probability of having a broken link.
}
\end{abstract}
\maketitle

\lettrine{I}n the field of quantum walks many results have been recently obtained, from both a theoretical \cite{KempeRev,VenegasReview,Kendon_ReviewDecoherence,Ref2,Ref9} and an experimental \cite{KarskiScience,SchmitzPRL,PhysRevLett.104.100503,PhysRevLett.104.153602,SchreiberScience,DelayedChoice, Ref1,Ref3,Ref4} points of view. Being the quantum analogue of the classical random walk, the quantum walk \cite{PhysRevA.48.1687} is a simple yet interesting model that can be exploited in several ways. It has been proved that quantum walks can be used to implement quantum search algorithms \cite{PhysRevA.67.052307}, and they can be considered as a universal computational primitive \cite{PhysRevLett.102.180501,PhysRevA.81.042330}. They have applications in the simulation of biological processes \cite{Photosyntetic}, and an equivalence with quantum lattice gas models \cite{Feynman} has been proposed. It is worth to keep in mind that the scenario with only a single walker is however classically simulable, for example using coherent light instead of a single-photon state in an optical implementation \cite{MultiWalker,PhysRevA.69.012310}. In order to have a true quantum behaviour, that could possibly lead to  computational protocols achieving an exponential speedup over their classical counterparts, the complexity introduced by the presence of more than one walker has to be considered (see for instance the short discussion in \cite{2010arXiv1010.4608R}). 
From this perspective, the study of the emerging features of multi-particle quantum walks, together with an analysis of their resilience to non-perfect conditions (e.g. noisy or disordered environments), is of paramount importance.  
For example, it is well known that a quantum particle in a static disordered media cannot evolve arbitrarily far from its starting point, a phenomenon known as ``Anderson localisation'' \cite{Anderson_Loc}. Such behaviour has been observed also for particles performing a quantum walk, and their localisation length does not considerably change even in the presence of interactions \cite{Ref5,Ref8,Ref13}.

In this manuscript we focus on discrete-time quantum walks of non-interacting particles. This model is based on the repeated application of a single-step evolution operator,  applied independently on both walkers. Such multi-particle scenario has been studied when a fixed unitary evolution is considered \cite{PhysRevA.75.032351,PhysRevA.74.042304,MeetingProb,MultiWalker,PhysRevA.81.052313,2009arXiv0901.3946V,PhysRevLett.108.010502}. 
Since this could not always be the case, it is interesting to study what happens with less stringent requirements. In the following we will consider the possibility of having missing links in the underlying structure (percolation graphs), modelling for example physical situations where one does not have perfect control over the lattice in which the walk takes place.
Up to now, this scenario has been investigated by several authors only in the single-particle case, approaching the problem from different perspectives: e.g. studying decoherence \cite{RomanelliPercolation} or disordered systems \cite{Kendon_Percolation}, modelling transport phenomena \cite{chandrashekar2014quantum}, or describing asymptotic behaviours \cite{Kollar}. 
 
The purpose of this paper is to understand how the possible absence of links (due, for instance, to a non-perfect experimental setup) could influence the evolution of multiple walkers, a fundamental step toward a complete exploitation of the quantum-walk model. In order to do so, we study the walk of two non-interacting particles, using as underlying lattice a simple line with bonds randomly missing with time.  More precisely, on the one hand we want to see to what extent signatures of the presence of multiple walkers survive when percolation is considered, while on the other hand we are searching for possible advantages arising when dealing with two particles instead of one.
To achieve the first goal, we qualitatively describe how different quantities depend upon the probability of having missing bonds and on the frequency with which the lattice changes. In particular, we will consider the final average distance between the walkers and the probability of finding both particles at the same location or exactly in their initial position, showing how their values strongly depend on the considered initial state (potentially entangled) even in noisy regimes with many broken links.
Concerning the advantages of having a second walker we show that, although the spread of the walk is not considerably altered by the addition of a particle, there is a procedure to estimate the probability of having a missing link whose performance is often improved by its presence. 
Furthermore, from our study it emerges that a signature of enhanced bunching between the walkers in the output probability distribution does not necessarily imply the presence of entangled states symmetric under particle-exchange. Since the symmetry condition is required by quantum indistinguishability, this observation is a warning against the possibility of gathering information about indistinguishability from the mere presence of bunching effects, at least in all those situations that deal with a random component in the evolution by considering averaged quantities. This is an important feature that has to be kept in mind also when studying more general many-walker scenarios in future research.

Let us point out that similar studies in more complex geometries, as higher-dimensional lattices \cite{HighDimensions} or graphs \cite{aharonov2001quantum}, could be of interest. However, being more computationally demanding, this is left as a possibility for future investigations. In the remainder of this introduction we will review some basic facts about quantum walks and imperfect lattices.


\vskip0.5cm
\noindent
{\bf Single-particle discrete-time quantum walk without percolation.}
Let us briefly review the concept of discrete-time quantum walk on the line. This can be thought as the quantum counterpart of a classical random walk, where at each step the walker can move left or right with a certain fixed probability. In this scenario, each step can be considered as composed of two parts: a (possibly biased) coin toss and a shift in the direction associated with the random output. Initially introduced in \cite{PhysRevA.48.1687}, for a more detailed discussion on the quantum-walk model we refer to \cite{KempeRev,VenegasReview}.

Quantum mechanically the position of the walker is described by orthonormal quantum states $\{\ket{i}\}_i$, that span the Hilbert space $\mathcal{H}_P$, while the coin can be taken into account by adding a two-dimensional Hilbert space $\mathcal{H}_C$, so that the global system can be described in $\mathcal{H}_P\otimes\mathcal{H}_C$. The coin toss can be realised by means of a fixed unitary operation applied to the $C$ component of the system. Throughout this manuscript, we will use for this purpose the Hadamard matrix
\begin{equation}\label{eq: HadamardCoin}
H_C=\frac{1}{\sqrt{2}}\left(\begin{array}{cc}
1 & 1\\
1 & -1
\end{array}\right),
\end{equation} 
expressed in the standard basis of the coin $\{\up, \down\}$, whose states correspond to the classical idea of ``head'' and ``tail''. This matrix represents the simplest choice for an unbiased walk, with no direction privileged by the unitary evolution. Although it has been shown that the Hadamard coin leads to all possible quantum walks if different starting states are considered \cite{HadamardGenerality}, such equivalence is not guaranteed when percolation is introduced. Our choice has hence to be considered as motivated by the sake of simplicity. 
The second part of a single step corresponds to the shift of the walker in the lattice, toward a direction depending on the coin output. Formally, this can be written as the following operator acting non-trivially on the whole space $\mathcal{H}_P\otimes\mathcal{H}_C$:
\begin{equation} \label{def: basic shift operator}
S=\sum_i \Big(\ketbra{i+1}{i}\otimes\up\bup + \ketbra{i-1}{i}\otimes\down\bdown\Big),
\end{equation}
where we wrote the position component in front of the coin component. We will keep the same order throughout the manuscript.

\begin{figure}
	\centering
\subfloat[\label{fig: single particle qwalk}]{\includegraphics[scale=0.45]{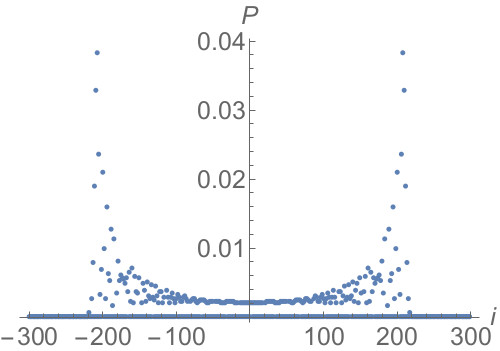}}\hskip0.25cm
\subfloat[]{\includegraphics[scale=0.45]{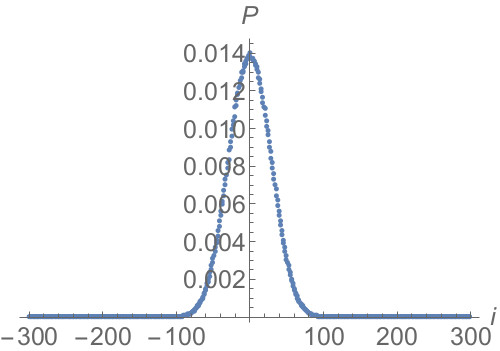}}\hskip0.25cm
\subfloat[\label{fig: SP-statical}]{\includegraphics[scale=0.45]{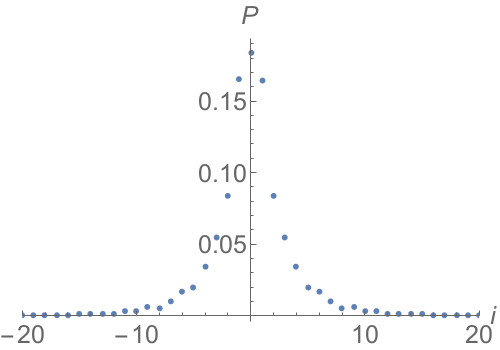}}
\caption{Average probability distribution for a quantum walk on the line after $N=300$ steps, without missing bonds $(a)$ and with dynamical $(b)$ or statical $(c)$ percolation (different ranges for the position values were chosen in order to increase the readibility of the plots). In order to have a symmetric distribution we started from the origin with a coin state given by Eq. \eqref{def: symm single particle states}. In the last two plots, we considered lattices characterised by a percolation parameter $p=0.75$ and we averaged over $10^5$ different outputs. \label{fig: standard q walk percolation}}
\end{figure}

A single step of the evolution can therefore be obtained by applying the operator 
\begin{equation}\label{def: single particle U}
U\equiv S\;(\Id_P\otimes H_C)
\end{equation}
to the vector $\ket{\psi_{in}} \in \mathcal H_P\otimes\mathcal H_C$,
representing the initial state of the system. In order to obtain a different result with respect to a classical random walk, a measurement is performed in the position degree of freedom only after a certain number of steps $N$. The probability distribution characterising the result will be 
\begin{equation} \label{eq: output single walker}
\mathcal{P}_1\dep{\ket{\psi_{in}}}(i)=\Tr\left[(\ketbras{i}{i}{}\otimes\Id_C)\, U^{(N)}\ketbra{\psi_{in}}{\psi_{in}}U^{(N)\dagger}\right],
\end{equation}
where the subscript $``1"$ reminds us that it represents the probability associated with a single walker. Without percolation, the $N$-step evolution operator $U^{(N)}$ is simply obtained by exponentiating $N$ times the single-step operator $U$ defined in Eq. \eqref{def: single particle U}.
After the measurement, that instance of quantum walk will be considered as concluded with output given by \eqref{eq: output single walker}.
It is well known that the average distance travelled by a classical random walker scales with the number $N$ of steps as $\sqrt{N}$, while for a quantum walker it scales linearly with it \cite{VenegasReview}. A typical plot for the probability distribution of a quantum walk can be observed in Fig. \ref{fig: single particle qwalk}.

Notice that all coefficients appearing in the evolution operators \eqref{eq: HadamardCoin} and \eqref{def: basic shift operator} are real. This implies that the real and imaginary parts of the initial state (expanded in the basis $\{\ket{i}\up$, $\ket{i}\down\}$) evolve independently. Therefore, a  conjugation of the state amplitudes does not change the final probability distribution, a fact that will be used later on.

\vskip0.5cm
\noindent
{\bf Percolation in 1D quantum walks.}
The situation previously described changes if we consider a lattice whose bonds can be randomly missing with time. Typically, in a percolation graph each bond is actually present only with a fixed probability $0\leq p \leq 1$, called ``percolation parameter'', and the structure is periodically changed respecting such probabilistic constraint. For more information on percolation lattices and their applications we refer to \cite{Percolation_Book,Percolation_Review}. Depending on the frequency of the changes, there can be two extreme regimes of percolation: \emph{statical} (different lattices for different walks) or \emph{dynamical} (different lattice at each step). The first scenario can be associated with a time scale for the lattice-imperfections evolution that is long with respect to the typical time required to finish a single walk, but small with respect to the average time needed between different runs of the apparatus. On the other hand, the regime of dynamical percolation appears when the imperfections in the lattice evolve so quickly that the missing bonds can change position even between consecutive steps of the same walk. In the following both these extreme cases will be considered.

In the presence of percolation, the evolution outlined in Eq. \eqref{def: single particle U} must be changed to avoid crossing of missing bonds. To achieve this the step operator applied on the $i$-th site is chosen as one of the following:
\begin{itemize}
			\item $S=\ketbra{i+1}{i}\otimes\up\bup + \ketbra{i-1}{i}\otimes\down\bdown$,
			\item $S_+ = \ketbra{i}{i}\otimes\down\bup + \ketbra{i-1}{i}\otimes\down\bdown$,
			\item $S_- = \ketbra{i+1}{i}\otimes\up\bup + \ketbra{i}{i}\otimes\up\bdown$,
			\item $S_\pm = \ketbra{i}{i}\otimes\down\bup + \ketbra{i}{i}\otimes\up\bdown$, 
\end{itemize}
respectively when both neighbouring bonds are present, when the following or the previous one is missing, or when they are both missing.

Every time the lattice changes, the missing-bonds positions are chosen randomly, so that eventually the output probability distribution has to be averaged over many different lattice sequences. As a result, the ballistic spreading typical of a quantum walk is lost, and the final probability distribution approaches that of a classical random walk \cite{Kendon_Percolation,RomanelliPercolation}, similarly to what happens with other decoherence models \cite{Kendon_ReviewDecoherence,Werner_Decoherence,Werner_CoinDecoherence}. An example can be seen in Fig. \ref{fig: standard q walk percolation}, where the outcomes of a $300$-step walk are shown in different percolation regimes. While in the standard quantum walk only odd (or even) positions can be occupied, there is no such constraint when the aforementioned strategy of time evolution with missing bonds is adopted. Notice also how in the dynamical case the spread of the Gaussian-like shaped distribution of the walker is much larger than the one of the statical output probability, which is narrowly peaked. This is a typical feature emerging in the two percolation regimes after a certain number of steps given approximately by $\overline L(p)/2$, where $\overline L (p) = p/(1-p)$ is the average length of a connected segment of the line, representing the length scale over which percolation becomes relevant. After this point the walker evolves with a diffusive behaviour in a dynamical percolation regime \cite{RomanelliPercolation}, while in the statical one it remains always localised within a segment: the subsequent average leads to the narrowly peaked distribution of Fig. \ref{fig: SP-statical}, independently of the number of steps.


\vskip0.5cm
\noindent
{\bf Non-interacting particles.}
In this last introductory section we present the formalism adopted throughout the paper to describe two non-interacting walkers. 
Using the same model of \cite{PhysRevA.74.042304}, the global Hilbert space of the system is $\mathcal{H}_{12}=\mathcal{H}_1\otimes\mathcal{H}_2$, where each $\mathcal H_k$ can be expanded as $\mathcal{H}_k = \mathcal{H}_{P_k}\otimes\mathcal{H}_{C_k}$. 
The single-step operator characterising the evolution is chosen to be the single-particle operator $U$ defined in Eq. \eqref{def: single particle U} (or modified close to the missing bonds according to the previous rule) applied independently to both systems: 
\begin{equation} \label{def: two walker evolution}
\mathcal{U}_{12}=U_1\otimes U_2,
\end{equation} 
where the label represents the Hilbert space of application.
This choice guarantees that both particles evolve independently, without any kind of interaction. If the evolution takes place on a perfect lattice,  $\mathcal{U}_{12}^{(N)}$ is obtained from \eqref{def: two walker evolution} via exponentiation, i.e., $\mathcal{U}_{12}^{(N)} = \mathcal{U}_{12}^{N}$. This changes in the presence of dynamical percolation, because every step is characterised by a different step operator (associated with a different lattice). 
	
In particular, in this manuscript we want to focus on indistinguishable particles, satisfying boson-like or fermion-like statistics or just being classically impossible to discriminate when detected at the end of the walk. 
A comparison of the corresponding results will allow us to distinguish the features due to the quantum nature of the particles from those coming from the averaging process.
Two single-particle states $\ket{\psi_1}$ and $\ket{\psi_2}$ can be combined to describe two bosons or two fermions by taking as joint initial state their opportune symmetrisation:
\begin{equation} \label{def: symmetrized state}
\ket{\text{Sym}_\pm(\psi_1,\psi_2)}_{12}=
\frac{\ket{\psi_1}_1\otimes\ket{\psi_2}_2\pm\ket{\psi_2}_1\otimes\ket{\psi_1}_2}{\sqrt{2\left(1\pm\big|\braket{\psi_1}{\psi_2}\big|^2\right)}},
\end{equation}
where the indeterminate form $\ket{\text{Sym}_-(\psi,\psi)}_{12}$ is set to $0$ by definition.
The case of classically indistinguishable particles, instead, can be modelled by considering an initial separable state $\ket{\psi_1}_1\otimes\ket{\psi_2}_2$ and by changing the measurement stage so as to make the final probability symmetric under particle-exchange. This is done by considering a measurement involving the set of projectors 
\begin{equation}\label{def: symmetrized projector}
\varPi_{ij}=\varPi_{ji}=\frac{1}{2}\left(\ketbra{ij}{ij}_{P_1,P_2}+\ketbra{ji}{ji}_{P_1,P_2}\right),
\end{equation}
that leads to a probability distribution
\begin{equation} \label{eq: output two walker}
\mathcal{P}_2\dep{\ket{\psi_{in}}}(i,j)=\Tr\left[\varPi_{ij}\; \mathcal{U}_{12}^{(N)}\ketbra{\psi_{in}}{\psi_{in}}\mathcal{U}_{12}^{(N)\dagger }\right].
\end{equation}
Notice that when dealing with bosons or fermions the addition of the symmetrised term in the projector does not affect at all the output probability.
With definition \eqref{eq: output two walker}, the probability of finding one particle in position $i$ and the other one in $j\neq i$ is given by $2 \mathcal{P}_2\left[\ket{\psi_{in}}\right](i,j)$, while the probability of finding them at the same location $i$ is just $\mathcal{P}_2\left[\ket{\psi_{in}}\right](i,i)$. Such description is chosen in order to obtain an output distribution properly normalised to $1$ on the whole plane $(i,j)$. 
Given two single-particle states $\ket{\psi_k}$, the output probability distributions obtained from \eqref{eq: output two walker} in the three cases (bosons, fermions or classically indistinguishable particles) can be labelled respectively by $\mathcal P_2^{(\pm)}$ and $\mathcal P_2^{(cl)}$. The dependence upon the states $\ket{\psi_k}$, $k=1,2,$ will be dropped most of the time to simplify the notation, when there is no possibility of confusion. While the first two distributions cannot generally be expressed in terms of single-particle outputs \eqref{eq: output single walker}, for two classically indistinguishable particles the following equality holds:
\begin{equation}
\label{eq: equivalence cl - single particle}
P_2^{(cl)}(i,j)=\frac{1}{2}P_1\dep{\ket{\psi_1}}(i)P_1\dep{\ket{\psi_2}}(j) + \frac{1}{2}P_1\dep{\ket{\psi_2}}(i)P_1\dep{\ket{\psi_1}}(j),
\end{equation}
as can be easily proved by using the factorised structure of both evolution \eqref{def: two walker evolution} and projector \eqref{def: symmetrized projector}. 

\begin{figure}[t]
	\centering
	\subfloat[\label{fig: perfect bos}]{\includegraphics[scale=0.45]{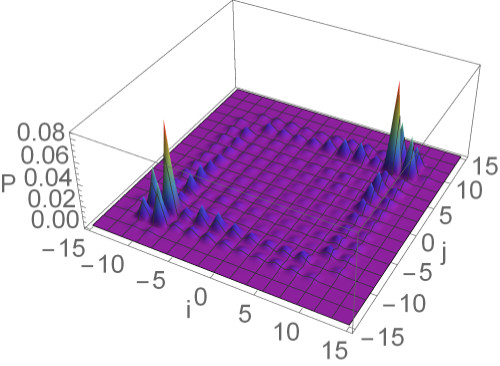}}\hskip0.25cm
	\subfloat[\label{fig: perfect ferm}]{\includegraphics[scale=0.45]{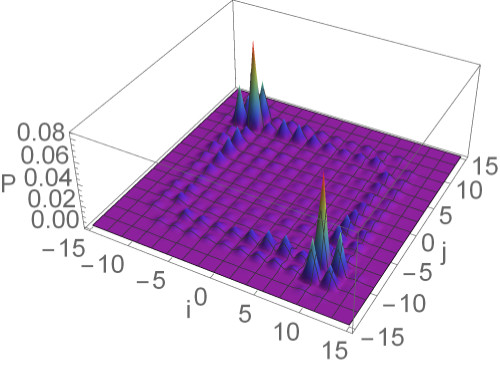}}\hskip0.25cm
	\subfloat[\label{fig: 2c}]{\includegraphics[scale=0.45]{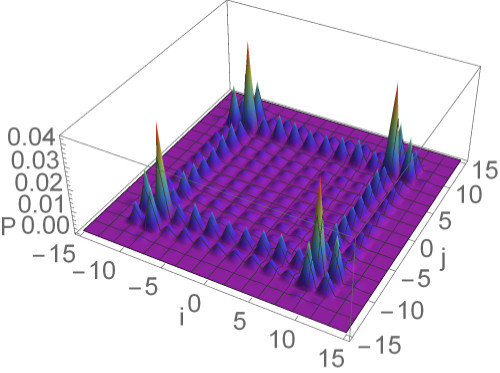}}
	\caption{Output probabilities for a 15-step quantum walk of two bosons, fermions or classically indistinguishable particles [respectively in $(a)$, $(b)$ and $(c)$], starting from the origin with the coin states given in Eq. \eqref{def: entangled states}.\label{fig: 2 walkers NoP}}
\end{figure}

It is worth noticing that not every entangled state in $\mathcal{H}_{12}$ can be expressed in a symmetrised form as in \eqref{def: symmetrized state}. An example of bipartite state left out by that particular structure is 
\begin{equation}
\ket{\psi^\text{(NotSym)}}=\frac{\ket{ij} + \ket{ji}}{\sqrt{2}}\otimes\frac{\updown + \downup}{\sqrt{2}},
\end{equation}
when $i\neq j$. In the following, however, we will restrict our investigation to states of a symmetrised form as in \eqref{def: symmetrized state}, or to separable states $\ket{\psi_1}\otimes\ket{\psi_2}$, because we need a definition of $\ket{\psi_k}$, $k=1,2$, to compare $P_2^{(\pm)}$ with $P_2^{(cl)}$. 
Moreover, in order to maximise the multi-particle effects in our interaction-free model, we will always consider states with a localised common initial position: the origin. In addition to this, to avoid asymmetrical spreading, we will focus mainly on combinations of the two single-particle coin states
\begin{equation}\label{def: symm single particle states}
\ket{\varphi_{\pm}}=\frac{\up \pm i \down}{\sqrt{2}},
\end{equation}
that lead to the symmetric evolution shown in Fig. \ref{fig: standard q walk percolation}.
With these conditions, the corresponding coin states for bosons, fermions and classically indistinguishable particles read
\begin{equation}\label{def: entangled states}
\ket{\phi_+}=\frac{\upp + \downn}{\sqrt{2}},\quad
\ket{\psi_-}=\frac{\updown - \downup}{\sqrt{2}},\quad
\ket{\psi_S}=\ket{\varphi_+}\otimes\ket{\varphi_-}.
\end{equation}
Notice that, with the initial state $\ket{\psi_S}$, relation \eqref{eq: equivalence cl - single particle} further simplifies: being $\ket{\varphi_+}$ and $\ket{\varphi_-}$ in Eq. \eqref{def: symm single particle states} the complex conjugate of each other, their single-particle output distributions are the same:
\begin{equation}
 \mathcal{P}_1[\ket{\varphi_+}] = \mathcal{P}_1[\ket{\varphi_-}] = \mathcal{P}_1, 
\end{equation}  
so that the joint probability $\mathcal{P}_2^{(cl)}(i,j)$ becomes
\begin{equation}\label{eq: equality symm NoP}
\mathcal{P}_2^{(cl)}(i,j) = \mathcal{P}_1(i)\mathcal{P}_1(j). 
\end{equation}
Examples of the probability distributions obtained by using the initial states defined in Eq. \eqref{def: entangled states} can be observed in Fig. \ref{fig: 2 walkers NoP}, where typical bunching and anti-bunching behaviours can be noticed (see \cite{PhysRevLett.108.010502} for an experimental realisation).

\vskip0.5cm
\noindent
{\large {\bf Results}}\\
\noindent
We can now finally address the problem of considering the quantum walk of two indistinguishable and non-interacting particles in a one-dimensional percolation lattice, where every run of the walk (in the statical case) or even every step (in a dynamical regime) takes place on a different underlying structure, randomly generated as described above. Considering a number $A\in \mathbb N$ of quantum walk realisations, the average output probability distribution reads
\begin{equation}\label{def: aver 2 particles prob}
	 \overline{P}_2(i,j)=\frac{1}{A}\sum_{a=1}^A P^{(a)}_2(i,j),
\end{equation} 
where index $a$ runs from $1$ to $A$. The evaluation of such quantity for different input states will be the basic tool adopted to characterise and study the behaviour of the walk.
Usually the plots in this manuscript are obtained by averaging over a number of different outputs varying from $A=2000$ to $A=5000$, depending on the precision required and on the complexity of the simulation. The obtained results do not qualitatively change in a significant way in that range. Moreover, they do not strongly depend upon the number of steps considered, that will typically be $N=15$ (we will explicitly point out the features that depend upon the parity of $N$). Therefore, we can use the obtained information to draw conclusions about the general qualitative behaviour of the walk.
Our discussion will include three sections, addressing three different issues.

At first, we want to qualitatively describe the average output distribution obtained in different percolation regimes considering the symmetric input states of Eq. \eqref{def: entangled states}. The goal of this analysis is to gain a better intuitive understanding of the situation, pointing out the features that are due to the quantum interference between the particles (that arises for entangled initial coin states, associated with bosonic or fermionic statistics) and those effects whose presence is rooted in the average appearing in Eq. \eqref{def: aver 2 particles prob}.
In particular, we show how a signature of enhanced bunching between the walkers in the output distribution does not necessarily imply the presence of entangled states symmetric under particle-exchange. This is therefore a warning against the possibility of gathering information about quantum indistinguishability from the mere presence of bunching effects.

In the second part of our discussion, we study how some quantities of interest depend upon the parameters of the problem (percolation regimes, probability of missing bonds and input states). In particular we focus on the average distance between the positions in which the particles are detected ($D$) \cite{PhysRevA.74.042304} and on the probability of finding them either in the same location ($M$), also called ``meeting probability'' \cite{MeetingProb,BookPercolation}, or exactly in the origin ($C$). On the one hand, this aims at capturing some particular aspects of the output by means of a few parameters that can be more easily interpreted and manipulated with respect to the whole probability distribution. Being the whole distribution impossible to plot for higher dimensions of the lattice or for a larger number of particles, in future research this kind of approach will become more and more  common. From this perspective, our analysis aims at highlighting the qualitative trends that appear in the aforementioned quantities in different scenarios of interest, whose identification will ease the interpretation of future, more complex, results. On the other hand, we are interested to see if the behaviour of these quantities strongly depends upon the considered input state [chosen among \eqref{def: entangled states}], implying the relevance that inter-particle quantum interference maintains even in very noisy regimes. This fact, observed in particular for the meeting probability $M$, can be interpreted as a signature of resilience of multi-particle effects to structural imperfections.

Finally, in the third and last section, we search for possible advantages arising when considering a multi-particle quantum walk with respect to its single-particle counterpart. At first we consider the single-particle spread, showing that it cannot be improved by the addition of another particle. However, inspired by the qualitative behaviour of $C$ that was previously found, we show how such quantity can be used in a procedure to estimate the percolation parameter (probability for each link to be present) whose performance can be improved by the presence of a second walker. 

\begin{figure}
	\centering
	\subfloat[\label{fig: perc dyn}]{\includegraphics[scale=0.45]{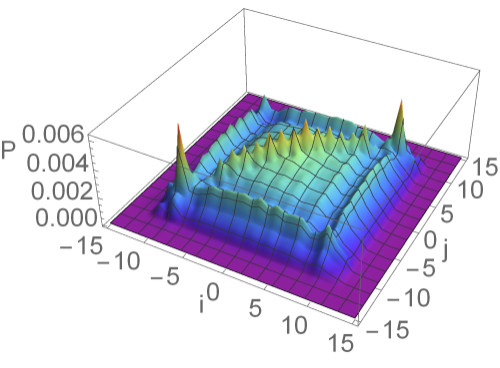}}\hskip0.5cm
	\subfloat[\label{fig: perc stat}]{\includegraphics[scale=0.45]{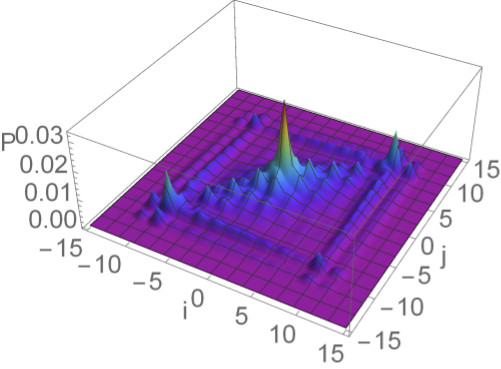}}
	\caption{$\overline{\mathcal{P}}^{(cl)}_2(i,j)$ for a quantum walk of two walkers starting from the same position with coin state $\ket{\psi_S}$ [see Eq. \eqref{def: entangled states}], measured after $15$ steps. Panels $(a)$ and $(b)$ correspond respectively to a dynamical and statical percolation regime, with $p=0.9$ and a number of averages $A=1000$. \label{fig: dyn-stat of a symmetric state}}
\end{figure}

\vskip0.5cm
\noindent
{\bf A. Analysis of bunching events.}
The typical behaviour of the average output distribution $\overline P_2^{(cl)}$ for classically indistinguishable particles, defined by evaluating Eq. \eqref{def: aver 2 particles prob} with the separable input state $\ket{\psi_S}$ of Eq. \eqref{def: entangled states}, is plotted in Fig. \ref{fig: dyn-stat of a symmetric state} for dynamical and statical percolation. As can be appreciated, the two results are qualitatively dissimilar. In particular, we can notice how the distribution is more spread out over the allowed region in the dynamical case, while in the statical regime it is concentrated in a few particular points, mainly along the diagonal. This result can be interpreted by keeping in mind that in the first case the walkers spread in a diffusive way, while in every run of a statical regime they are localised on a small segment around the origin. The final distribution is then obtained by averaging the output of each run, whose detection probability is similar to Fig. \ref{fig: 2c}. This leads to the diagonal peaks visible in Fig. \ref{fig: perc stat}, where the corresponding off-diagonal peaks of Fig. \ref{fig: 2c} disappeared because it is very unlikely to have a connected segment that extends far away from the origin in both directions.
Similar considerations could be made for entangled coin states, representing boson-like and fermionic-like statistics, if instead of Fig. \ref{fig: 2c} we consider as underlying coherent evolutions for example those given in Fig.s \ref{fig: perfect bos} and \ref{fig: perfect ferm}.
From this analysis it emerges that the so called ``bunching events'', associated with the detection of two particles at the same location, depend on the interplay between statistical average and effects of quantum interference between multiple walkers. In the following we want to formalise these considerations.

\begin{figure*}
	\centering
	\subfloat[$\overline{\mathcal{P}}^{(+)}_2(i,j)$]{\includegraphics[scale=0.45]{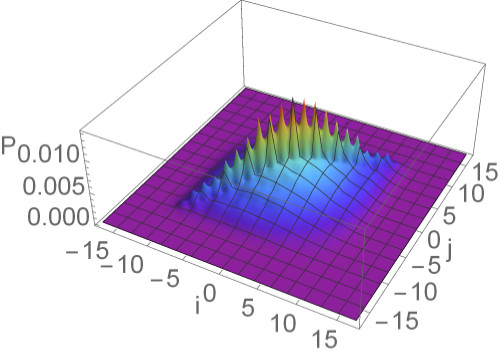}}\hskip0.3cm
	\subfloat[$\overline{\mathcal{P}}^{(-)}_2(i,j)$]{\includegraphics[scale=0.45]{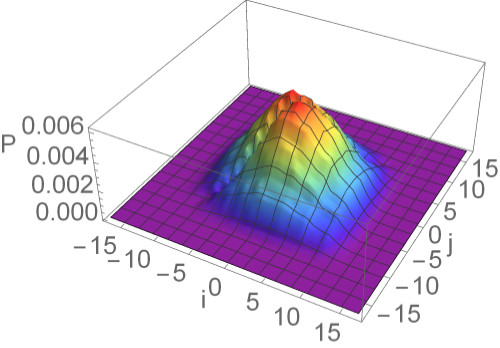}}\hskip0.3cm
	\subfloat[$\overline{\mathcal{P}}^{(cl)}_2(i,j)$]{\includegraphics[scale=0.45]{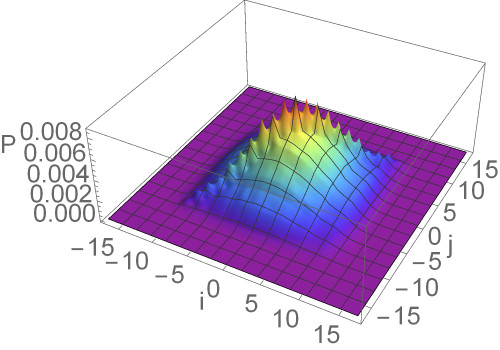}}\hskip0.3cm
	\vskip0.25cm
	\subfloat[Difference $(a) - (c)$]{\includegraphics[scale=0.45]{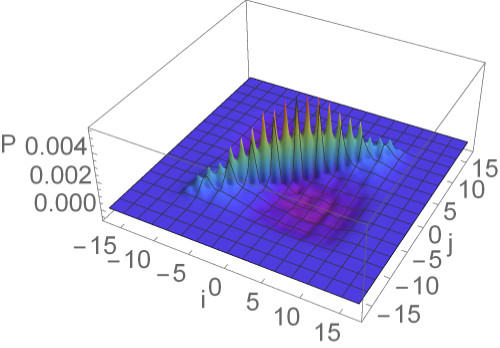}}\hskip0.5cm
	\subfloat[Difference $(b) - (c)$]{\includegraphics[scale=0.45]{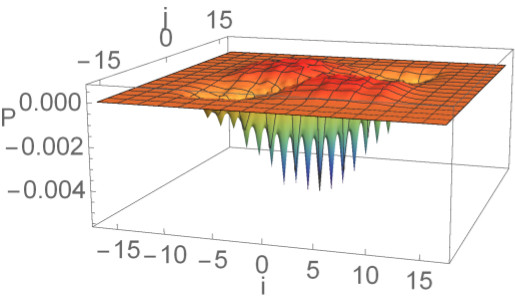}}
	\caption{Average output probability distributions for a two-particle 15-step quantum walk affected by dynamical percolation ($p=0.75$, $A=2000$). We considered boson-like $(a)$, fermion-like $(b)$, and classically indistinguishable $(c)$ combinations of two single-particle states $\ket{\varphi_\pm}$ that spread symmetrically from the origin. Panels $(d)$ and $(e)$ evidence the contribution to bunching events due to inter-particle quantum interference. \label{fig: bos/ferm peaks}}
\end{figure*}

In order to properly define a peak (or a valley) in an output distribution, we need a reference probability to use as comparison. The peaks (or valleys) due to inter-particle quantum interference can be evidenced by comparing the average probabilities $\overline{\mathcal{P}}_2^{(\pm)}$, associated with the symmetrisation of two single-particle input states  $\ket{\psi_1}$ and $\ket{\psi_2}$ [see Eq. \eqref{def: symmetrized state}], with the average probability $\overline{\mathcal{P}}_2^{(cl)}$, corresponding to classically indistinguishable particles starting from a separable input state obtained by taking the tensor product of the same $\ket{\psi_1}$ and $\ket{\psi_2}$. 
Indeed, for a single run of the quantum walk, whose evolution is labelled by $a$, we can expand the probability of finding both particles in the same location as
\begin{equation}
\mathcal{P}_2^{(\pm)(a)}(j,j)=\frac{\mathcal{P}^{(cl)(a)}_2(j,j)}{1\pm|\braket{\psi_1}{\psi_2}|^2} \pm \frac{\left|\bra{\psi_1}\mathcal{U}_{(a)}^{(N)\dagger }\ket{j}_P\bra{j}\mathcal{U}_{(a)}^{(N) }\ket{\psi_2}\right|^2}{1\pm|\braket{\psi_1}{\psi_2}|^2}.
\end{equation}
From this, it follows that at least when $\braket{\psi_1}{\psi_2}=0$ [which is the case in the main situation of interest here, where $\ket{\psi_{1,2}} = \ket{0}\ket{\varphi_{\pm}}$] the elements on the diagonal are larger (for bosons) or smaller (for fermions) than the diagonal elements in the corresponding probability for classically indistinguishable particles, as can be expected. Since this property holds for every index $a$, it is still true for the average probability distribution. This effect in a regime of dynamical percolation can be observed in Fig. \ref{fig: bos/ferm peaks}.

\begin{figure}
	\centering
		\subfloat[$\overline{\mathcal{P}}^{(cl)}_2(i,j)$]{\includegraphics[scale=0.45]{cla.jpg}}\hskip0.25cm
		\subfloat[$\overline{\mathcal{P}}_1(i) \; \overline{\mathcal{P}}_1(j)$]{\includegraphics[scale=0.45]{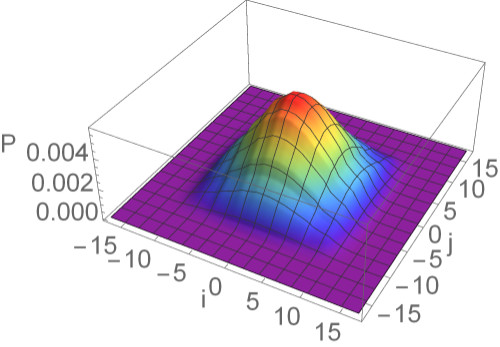}}\hskip0.25cm
		\subfloat[Difference $(a) - (b)$]{\includegraphics[scale=0.45]{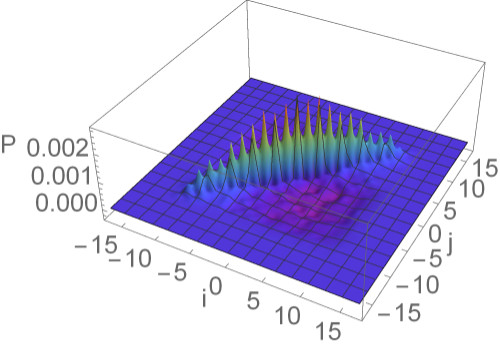}}
	\caption{$\overline{\mathcal{P}}^{(cl)}_2(i,j)$, $\overline{\mathcal{P}}_1(i)\overline{\mathcal{P}}_1(j)$, and their difference are plotted respectively in $(a)$, $(b)$, and $(c)$, for a quantum walk of $N=15$ steps starting from the origin with an initial coin state $\ket{\psi_S}$, defined in Eq. \eqref{def: entangled states}. A dynamical percolation regime with $p=0.75$ averaged over $A=2000$ different evolutions is considered. This figure shows how the statistical averaging procedure might enhance bunching events in the average probability distribution of a multi-particle quantum walk.\label{fig: equal tensor comparison}}
\end{figure}

On the other hand, the effect of the average can be understood by comparing $\overline{\mathcal{P}}^{(cl)}_{2}$ with the symmetrised product of the two single-particle average distributions $\overline{\mathcal{P}}_1\dep{\ket{\psi_i}}$. Surprisingly, in some situations the averaging procedure might induce an enhancement of the probability of bunching events, leaving a signature in the output distribution similar to the interference effect previously analysed. To see this, we can expand the probability $\mathcal{P}^{(a)}_1\dep{\ket{\psi_k}}$ around its mean value as
\begin{equation}
\mathcal{P}^{(a)}_1\dep{\ket{\psi_k}} = \overline{\mathcal{P}}_1\dep{\ket{\psi_k}} + \delta^{(a)}\dep{\ket{\psi_k}},
\end{equation}
so that, for every lattice site $j$, the differences $\delta^{(a)}\dep{\ket{\psi_k}}$ average to zero:
\begin{equation}
\frac{1}{A}\sum_{a=1}^A \delta^{(a)}\dep{\ket{\psi_k}}(j) =0,
\end{equation}
giving, for the average detection probability of two classically indistinguishable particles,
\begin{align}
\overline{P}_2^{(cl)}(i,j)&= \frac{1}{2}\left(\overline{P}\dep{(\ket{\psi_1})}(i)\;\overline{P}\dep{(\ket{\psi_2})}(j) + \overline{P}\dep{(\ket{\psi_1})}(j)\;\overline{P}\dep{(\ket{\psi_2})}(i)\right) \notag \\
&+ \frac{1}{2A}\sum^A_{a=1} \left(\delta^{(a)}\dep{(\ket{\psi_1})}(i)\;\delta^{(a)}\dep{(\ket{\psi_2})}(j) + \delta^{(a)}\dep{(\ket{\psi_1})}(j)\;\delta^{(a)}\dep{(\ket{\psi_2})}(i)\right). \label{eq: delta fluctuations}
\end{align}
Usually when $\ket{\psi_1}$ and $\ket{\psi_2}$ lead to different probability distributions, or when off-diagonal terms  $i\neq j$ are considered, the second line in \eqref{eq: delta fluctuations} has not a fixed sign, so for $A\gg1$ it does not contribute significantly. This is no longer true when we consider diagonal terms $i = j$ for states with the same final distribution (e.g. $\ket{0}\ket{\varphi_+}$ and $\ket{0}\ket{\varphi_-}$). In this case the second line in \eqref{eq: delta fluctuations} is always positive, being a sum of squared terms. These contributions hence add up to create the peaks over the first term, which in the example provided is just equal to the product $\overline{\mathcal{P}}_1(i)\overline{\mathcal{P}}_1(j)$ (see Eq. \eqref{eq: equality symm NoP} and Fig. \ref{fig: equal tensor comparison} for the resulting plots in a regime of dynamical percolation).

We expect that a similar behaviour can arise also with more complicated lattice structures affected by percolation. Therefore, whenever one deals statistically with such imperfections, it is useful to keep in mind that the mere averaging process could lead to an enhancement of the diagonal entries of the average probability distribution. Being aware of this fact can avoid the misinterpretation of such phenomenon as a bunching signature due to the quantum indistinguishability of the multiple walkers considered.
 
\vskip0.5cm
\noindent
{\bf B. Qualitative analysis of three significant quantities in the presence of percolation.}
In what follows, we want to see how a few quantities depend upon the adopted percolation parameter $p$ and the particle-exchange symmetry encoded in the input states, for both dynamical and statical percolation regimes. In particular, we will focus on two walkers starting from the same position with initial coin states $\ket{\phi_+}$, $\ket{\psi_-}$ or $\ket{\psi_S}$, defined in Eq. \eqref{def: entangled states}.
For every considered quantity $Q$ and for a set of equally spaced percolation parameters $p$, we will average the value obtained for a single lattice realisation $Q^{(a)}$ over $A=5000$ different configurations:
\begin{equation}\label{def: meet prob}
Q(p)=\frac{1}{A}\sum_{a=1}^A Q^{(a)}(p).
\end{equation}
The uncertainty associated with such average will be then given by the standard error
\begin{equation} \label{eq: err meet prob}
	\sigma_Q(p) = \frac{1}{\sqrt{A}}\sqrt{\frac{1}{A-1}\sum^A_{a=1}\left[Q^{(a)}(p)-Q(p)\right]^2}.
\end{equation}

\begin{figure}
	\centering
	\subfloat[]{\includegraphics[scale=0.6]{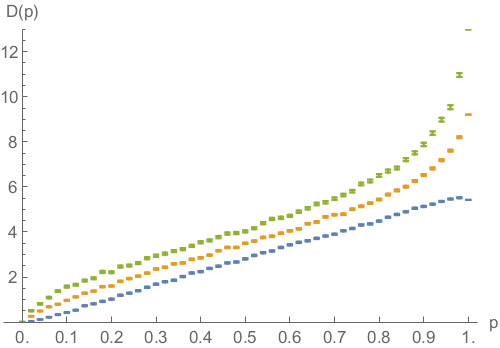}}\hskip0.5cm
	\subfloat[]{\includegraphics[scale=0.6]{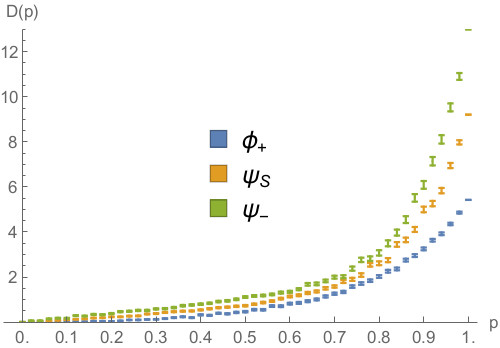}}
	\caption{Average output distance between the two particles as a function of the percolation parameter $p$, for a 15-step quantum walk starting from the origin with three different coin states [see Eq. \eqref{def: entangled states}]. $(a)$ and $(b)$ correspond respectively to dynamical and statical percolation regimes. The average is calculated over $A = 5000$ different lattice realisations and the errors are evaluated as the standard errors on the mean values obtained. \label{fig: output average distance}}
\end{figure}

\vskip0.5cm
\noindent
{\bf B1. Final average distance between the particles.}
A global property that can be analysed is the average distance between the positions where the two walkers are detected. Formally such quantity can be expressed as
\begin{equation}\label{def: averge distance formula}
D = \sum_{ij} |j-i|\;\mathcal{P}_2(i,j).
\end{equation} 
Its numerical evaluation for different percolation parameters $p$ is plotted in Fig. \ref{fig: output average distance}, for a 15-step quantum walk. As can be expected, for any value of the percolation parameter, this quantifier for fermions is larger than the one for bosons, with an intermediate behaviour for the separable coin $\ket{\psi_S}$. This result agrees with the discussion in \cite{PhysRevA.74.042304}, that considered this same quantity without percolation.

If we consider the dependence upon $p$ in the dynamical percolation regime, we can see that the average distance between bosons increases approximately in a linear way, with a small decrease just before $p=1$. This can be interpreted as a trade-off between two trends: on the one hand with less missing bonds the two walkers can move on a larger segment of the line, increasing the likelihood of finding them further away; on the other hand, if the two particles are less disturbed, their distribution approaches the one obtained without percolation (shown in Fig. \ref{fig: perfect bos}), characterised by a high degree of bunching. For fermions, instead, the same considerations add up increasing the average distance of separation, since they tend to show an anti-bunched distribution in the limit $p\to 1$ (Fig. \ref{fig: perfect ferm}).

If a comparison is made between the plots obtained with dynamical and statical percolation for a 15-step walk, one can notice that in the second regime the average distance between the particles is much smaller. This can be understood by thinking that with statical percolation in each run the particles are constrained to stay close to each other in a fixed region of space, whereas in the dynamical case they have always the chance of getting further away. The observed behaviour is hence typical once the number of steps is larger than the average length of the connected segment $\overline L$ available to the walkers in a regime of statical percolation \cite{RomanelliPercolation}.

\vskip0.5cm
\noindent
{\bf B2. Meeting probability.}
Another global quantity, easier to obtain with respect to the previous one as it involves a smaller number of measurements, is the overall probability for the two walkers to be on the same position at the end of the walk. Such quantity, also called meeting probability \cite{BookPercolation,MeetingProb}, corresponds to the sum of the diagonal terms in the output probability distribution $\mathcal{P}_2(i,j)$ in Eq. \eqref{eq: output two walker}: 
\begin{equation}
M = \sum_j \mathcal P_2(j,j).
\end{equation}

\begin{figure}
	\centering
	\subfloat[]{\includegraphics[scale=0.6]{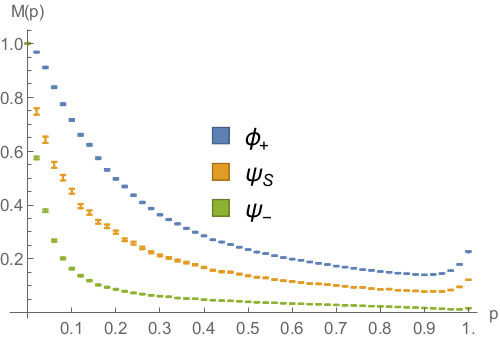}}\hskip0.5cm
	\subfloat[\label{fig: stat mp}]{\includegraphics[scale=0.6]{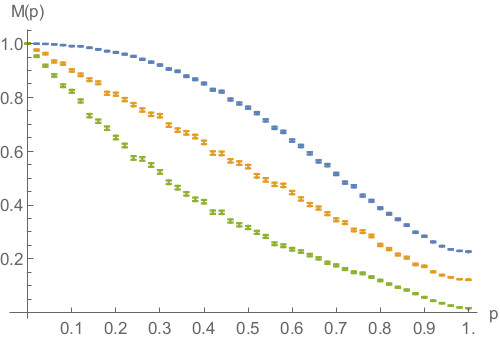}}
	\caption{Average meeting probability as a function of the percolation parameter $p$, for a 15-step quantum walk starting from the origin with three different coin states [see Eq. \eqref{def: entangled states}]. $(a)$ and $(b)$ correspond respectively to dynamical and statical percolation regimes. The average is calculated over $A = 5000$ different lattice realisations and the errors are evaluated as the standard errors on the mean values obtained. \label{fig: meeting prob}}
\end{figure}

The results obtained for a 15-step quantum walk are plotted in Fig. \ref{fig: meeting prob} for both dynamical and statical percolation. We can notice how in the first scenario the meeting probability decreases for $p\lesssim 0.92$ and slightly increases after this point, while in the statical case it is monotonically decreasing for all values of $p$. The minimum position in the dynamical case seems to depend on the number of steps, moving slowly toward $p=1$ when this is increased. However, even if this could look like a finite-size effect, we pushed simulations up to $80$ steps, and have strong numerical evidence that the non-monotonic behaviour would still be present even for a much larger number of steps. We can obtain an intuitive explanation for this fact by comparing the output of the walk in a perfect lattice (Fig. \ref{fig: 2 walkers NoP}) with the ``spreading'' effect that dynamical percolation has on the output probability distribution (see Fig. \ref{fig: perc dyn}). Starting from the ideal case of $p=1$, the addition of small amounts of imperfections makes the diagonal peaks of Fig. \ref{fig: 2 walkers NoP} spread, leading to a diminished meeting probability. This effect is much less important in a statical percolation regime, as can be appreciated in Fig. \ref{fig: perc stat}. This is because in every lattice realisation the walk is coherent on a small segment of the line around the origin, and the percolation only averages such small-scale coherent evolutions. The overall result is that in this case the meeting probability depends monotonically upon $p$ (see Fig. \ref{fig: stat mp}).

From Fig. \ref{fig: meeting prob} we can also see how the quantum statistics of the particles heavily affects the meeting probability, even for small values of $p$. This shows that a signature of bosonic or fermionic multi-particle effects survives also in a noisy regime where the quantum walk tends to its classical counterpart.

\vskip0.5cm
\noindent
{\bf B3. Probability of finding both walkers in the origin.}
Finally, we consider a ``local'' property, i.e. that can be obtained through a measurement in a single position: the probability of finding both walkers in their initial starting point. In the following we will label it as
\begin{equation}\label{eq: C(p)}
C = \mathcal{P}_2(0,0).
\end{equation}
In addition to being more easily obtained with respect to the previous quantities, this one differs also because it depends on the parity of the number of steps $N$ considered. Indeed, when $p=1$, $C$ is rigorously zero if $N$ is odd, but this could not be the case if $N$ is even. Being mostly interested in this quantity for its monotonic and regular behaviour, that can be exploited to estimate the percolation parameter $p$ (see next section), we will focus only on odd numbers of steps in order to have a quantity varying in the maximum range $[0,1]$.

\begin{figure}
	\centering
	\subfloat[]{\includegraphics[scale=0.6]{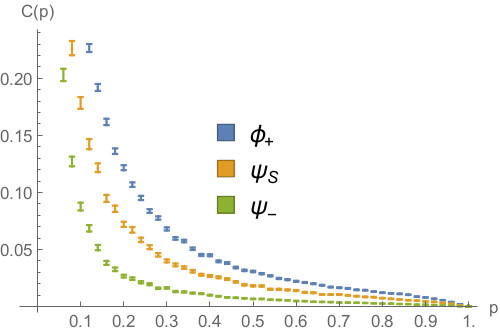}}\hskip0.5cm
	\subfloat[\label{fig: cent stat}]{\includegraphics[scale=0.6]{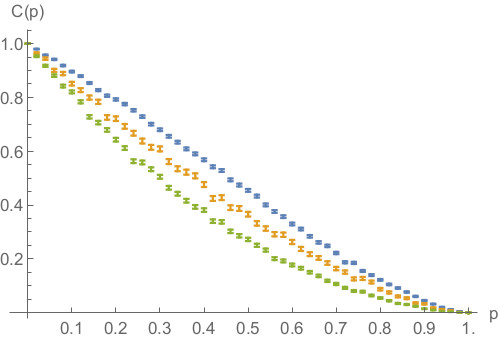}}
	\caption{Average probability of finding both walkers in the origin, as a function of the percolation parameter $p$, for a 15-step quantum walk starting from the origin with three different coin states [see Eq. \eqref{def: entangled states}]. $(a)$ and $(b)$ correspond respectively to dynamical and statical percolation regimes. The average is calculated over $A = 5000$ different lattice realisations and the errors are evaluated as the standard errors on the mean values obtained. \label{fig: meeting prob center}}
\end{figure}

Intuitively, $C$ will be close to $1$ for small values of $p$, since both walkers are constrained by the missing bonds to stay close to the starting point, and it decreases with increasing $p$ while the walk spreads further away, reaching zero for $p=1$ (for $N$ odd).  This behaviour is numerically confirmed by Fig. \ref{fig: meeting prob center}, that shows how $C$ depends upon $p$ for a 15-step quantum walk affected by dynamical or statical percolation. We can see how in the first regime $C(p)$ goes quickly to zero as soon as the imperfections are reduced. This can be explained with the diffusive behaviour of the walkers, that are allowed to quickly populate regions with $i\neq j$ in the $\mathcal P_2(i,j)$ plot (e.g. see Fig. \ref{fig: perc dyn}). The situation is different in the statical case, where the walkers move away from the origin in a more controlled way (e.g. see Fig. \ref{fig: perc stat}), based once again on the typical length of connected lattice around the origin. The resulting plot for $C(p)$ is monotonic with almost constant slope in the whole range $[0,1]$, as can be appreciated in Fig. \ref{fig: cent stat}.

\vskip0.5cm
\noindent
{\bf C. Search for multi-particle advantages.}
In this final section we want to discuss whether the addition of a second walker to the walk can lead to an immediate advantage in some task or protocol. As a first check we consider the possibility of having an improvement in the spread of the walk and show that this is not the case. For example, for every quantum statistics of the two particles involved there is always a single-particle state leading to a larger spread. We then consider a more specific task, namely the estimation of the percolation parameter affecting the lattice, proposing a two-walker procedure that in certain regimes shows an advantage over its single-particle counterpart.

\vskip0.5cm
\noindent
{\bf C1. Spread of the walk.}
The intuitive idea of spreading for a single walker, starting from a localised position $i_0$, can be well described by the quantity 
\begin{equation}\label{def: variance 1 walker}
V_1=\sum_i \left(i-i_0\right)^2 P_1(i),
\end{equation}
where the initial coin state is not explicitly written. Notice that, if the walker evolves symmetrically, such quantity corresponds to the variance of the output distribution. How can we generalise this expression to the two-walker case? Also here, we are mostly interested in letting both particles start from the same lattice position. Indeed, since in our model there are no interactions, this situation is the one that could maximise the multi-particle effects. For this reason, we will consider here only the case where both walkers start from the origin (with arbitrary coin states). With this hypothesis, the straightforward generalisation of \eqref{def: variance 1 walker} for a generic fixed unitary evolution labelled by $a$ (i.e. for a given sequence of lattices) is
\begin{equation}\label{def: variance 2W sym}
V_2^{(a)} = \frac{1}{2}\sum_{ij} \left( i^2 + j^2\right)\mathcal{P}^{(a)}_2(i,j).
\end{equation}
Once two single-particle coin states have been fixed, we can label with $V_2^{(\pm)(a)}$, $V_2^{(cl)(a)}$ the spread corresponding to the different statistics of the particles, obtained by using in \eqref{def: variance 2W sym} respectively $\mathcal P_2^{(\pm)(a)}$ or $\mathcal P_2^{(cl)(a)}$. In the same way, with an overbar we will label the value of the spread averaged over many different lattice sequences in a percolation regime. The overall factor $1/2$ in our definition \eqref{def: variance 2W sym} is introduced only to ease the comparison with the single-walker spread $V_1$ of Eq. \eqref{def: variance 1 walker}, because from Eq. \eqref{eq: average spread} it follows
	\begin{equation}\label{eq: average spread}
	V_2^{(cl)(a)} = \frac{V^{(a)}_1(c_1) + V^{(a)}_1(c_2)}{2}\leq \max_{k\in\{1,2\}}  V^{(a)}_1(c_k),
	\end{equation} 
where two generic single-particle coin states $\ket{c_1}$, $\ket{c_1}$ are considered. This is a simple consequence of the fact that two particles initialised in a separable state spread independently under the evolution in Eq. \eqref{def: two walker evolution}.  The fact that we cannot distinguish them in the measurement leads only to the average that appears in Eq. \eqref{eq: average spread}. 

One can now wonder whether the presence of a proper quantum statistics, introduced by the symmetrisation of the vector describing the state, can somehow alter this result. In order to show that the situation does not considerably change, we consider an initial state 
	\begin{equation}
	\ket{\psi^{(\pm)}_{in}} = \ket{00}\otimes  \ket{\text{Sym}_\pm(c_1,c_2)},
	\end{equation}
whose spread after $N$ steps (for a particular lattice sequence) can be written as 
	\begin{equation}\label{prop2 statement 1}
	V_2^{(\pm)(a)} = \sum_{k=1}^2 \lambda^{(\pm)}_k V^{(a)}_1\left(\tilde c^{(\pm)}_k\right).
	\end{equation}
	In this expression $0\leq\lambda^{(\pm)}_k\leq 1$ and $\ket{\tilde c^{(\pm)}_k} \in \mathcal H_{C_1}$ are the two eigenvalues and eigenvectors of the coin state
	\begin{equation}\label{def: lambda and c tilde}
	\bras{0}{P_1}  \Tr_2\left[\ketbra{\psi^{(\pm)}_{in}}{\psi^{(\pm)}_{in}}\right] \kets{0}{P_1},
	\end{equation} 
	obtained by tracing out the Hilbert space of the second particle. 
	Moreover, the odd combination $\ket{\text{Sym}_-(c_1,c_2)}$ of two generic coin states $\ket{c_1}, \ket{c_2} \in \mathcal{H}_C$, as well as the even one $\ket{\text{Sym}_+(c_1,c_2)}$ or separable one $\ket{c_1}\otimes\ket{c_2}$ of two orthogonal coin states  $\braket{c_1}{c_2} = 0$, yields the same spread as the two single-particle symmetrical-spreading states $\ket{\varphi_{\pm}}$ [defined in Eq. \eqref{def: symm single particle states}]:
	\begin{equation}\label{prop2 result}
	V_2^{(-)(a)}=V_2^{(+)(a)}\Big|_{\braket{c_1}{c_2} = 0} = V_2^{(cl)(a)}\Big|_{\braket{c_1}{c_2} = 0} = V^{(a)}_1(\varphi_{\pm}).
	\end{equation}			
The proof of these facts can be found in Methods.
Being $\ket{\varphi_\pm}$ orthogonal coin states leading to the same probability distribution, Eq.s \eqref{eq: average spread} and \eqref{prop2 result} imply 
\begin{equation}\label{eq: spread 1}
V^{(a)}_2(\phi_+)=V^{(a)}_2(\psi_-)=V^{(a)}_2(\psi_S)=V^{(a)}_1(\varphi_\pm).
\end{equation}
This means that there is no advantage, from the spread perspective, in considering two symmetrical-spreading particles instead of just one, no matter what the statistics over particle-exchange enforced in the input state by Eq. \eqref{def: symmetrized state} is.
From Eq. \eqref{prop2 result} it follows that only the bosonic statistics could a priori influence the spread, when two non-orthogonal single-particle coin states $\ket{c_1}$ and $\ket{c_2}$ are symmetrised. However, Eq. \eqref{prop2 statement 1} implies that there will always exist a single-particle state with a spread at least as large as the one found with two bosons, because
\begin{equation}\label{eq: two-part spread ineq}
V_2^{(+)(a)} = \sum_{k=1}^2 \lambda^{(+)}_k V^{(a)}_1\left(\tilde c^{(+)}_k\right) \leq \max_{k\in\{1,2\}} V^{(a)}_1\left(\tilde c^{(+)}_k\right).
\end{equation}
We emphasise that such states $\ket{\tilde c^{(+)}_k}$ could be different from both coin states $\ket{c_1}$ and $\ket{c_2}$ symmetrised at the beginning, and that the optimal $k$ could depend on the explicit lattice realisation labelled by ``$a$''.

All these results can be trivially extended to the spread averaged over many lattice configurations. In particular, for every initial coin states $\ket{c_1}, \ket{c_2}$ one has
\begin{equation}
\overline V_2^{(cl)} = \frac{\overline V_1(c_1) + \overline V_1(c_2)}{2} \leq \max_{k\in\{1,2\}} \overline V_1\left(c_k\right),
\end{equation} 
\begin{equation}\label{eq: averaged spread}
\overline V_2^{(+)} =  \sum_{k=1}^2 \lambda^{(+)}_k \overline V_1\left(\tilde c^{(+)}_k\right) \leq \max_{k\in\{1,2\}} \overline V_1\left(\tilde c^{(+)}_k\right), \quad\qquad \overline V^{(-)}_2= \overline V_1(\varphi_\pm),
\end{equation}
in terms of $\ket{\tilde c^{(+)}_k}$ and $\lambda_k^{(+)}$ defined in Eq. \eqref{def: lambda and c tilde}.

\vskip0.5cm
\noindent
{\bf C2. Estimation of the percolation parameter.}
Here we want to explicitly consider the possibility of using some of the quantities described in section $B$ to infer the value of the percolation parameter $p$ that characterises the lattice imperfections. In order to do so, let us consider a certain event $E$, happening at the measurement stage of the quantum walk: e.g. both particles at the same location, both in the origin, etc. The probability of such event will depend on the explicit lattice realisation, and we will write it as $P_E^{(\xi)}$, where $\xi$ runs over all the possible sequences of lattice configurations. Notice that here we are using $\xi$ instead of the previous $a$, to emphasise the fact that we are considering an effective lattice realisation and not of a simulation of it. $P_{E}^{(\xi)}$ has to be interpreted as the probability of the event considered, conditioned upon the $\xi$-th realisation of the lattice: $P_{E}^{(\xi)} = P(E|\xi)$. Therefore, the probability of $E$ can be written as
\begin{equation}
P_{E} = \sum_{\xi} P(E|\xi) \, P(\xi) = \sum_{\xi} P_{E}^{(\xi)} \, P(\xi).
\end{equation}

This quantity, although difficult to obtain analytically, can be numerically estimated. Indeed, for the event ``both particles in the same place'' (or ``both particles in the origin'') it would correspond to $M(p)$ \eqref{def: meet prob} [or $C(p)$ \eqref{eq: C(p)}] in the limit of many different sequences of lattice configurations considered ($A\to \infty$). In the following we will generically label this numerical estimate as $P^{(sim)}_{E}$, whose precision will depend on the number $A$ adopted [see the standard error in Eq. \eqref{eq: err meet prob}]. On the other hand, if one had access to an experimental setup of such quantum walk affected by percolation, after $n$ runs of the experiment $P_E$ could be estimated via the quantity
\begin{equation}
P_{E}^{(est)} = \frac{n_{E}}{n}, 
\end{equation}
where $n_{E}$ is the number of times in which $E$ happened. This estimator is a stochastic variable, characterised by expectation value and variance given by
\begin{equation}\label{eq: Var PE}
EV\left[P_{E}^{(est)}\right] = P_{E}, \qquad \text{VAR}\left[P_{E}^{(est)}\right] = \frac{P_{E}(1-P_{E})}{n}.
\end{equation} 
This can be shown by considering $P_{E}^{(est)}$ as the average of $n$ independent and identically distributed variables, taking the value $1$ with probability $P_{E}$ (when $E$ happens in that single run) and the value $0$ otherwise.

Our goal is to estimate the value of $p$ through a measurement of  $P_{E}^{(est)}$, that estimates $P_{E}$ with an error $\text{VAR}\left[P_{E}^{(est)}\right]^{1/2}$. Exploiting the linear propagation of errors and the explicit expression of the variance given in Eq. \eqref{eq: Var PE}, we can write the uncertainty in the estimation of $p$ as
\begin{equation}
\delta p = \frac{\sqrt{\text{VAR}\left[P_{E}^{(est)}\right]}}{\left|\frac{\partial P_{E}}{\partial p}\right|} = \frac{\sqrt{P_{E}(1-P_{E})}}{\sqrt{n} \left|\frac{\partial P_{E}}{\partial p}\right|},
\end{equation}
so that an upper bound $\delta p \leq \epsilon$ yields a lower bound on the required number of runs $n \geq n_{min}^{(\epsilon)}(p)$:
\begin{equation}\label{eq: Nmin formula}
n\geq n_{min}^{(\epsilon)}(p) = \frac{P_{E}(1-P_{E})}{\epsilon^2 \left|\frac{\partial P_{E}}{\partial p}\right|^2}.
\end{equation}
The right-hand side can be estimated by using $P^{(sim)}_{E}$, numerically obtained by considering a total number of $A \gg 1$ sequences of lattices. In particular we can approximate 
\begin{align}
P_{E}(1-P_{E}) &\simeq 	P_{E}^{(sim)}(1-P^{(sim)}_{E}), \label{eq: biased estimator}\\
\frac{\partial P_{E}}{\partial p} &\simeq \frac{\partial P^{(sim)}_{E}}{\partial p},\label{eq: fit comment}
\end{align}
where the second term is obtained from a polynomial fit of the numerical samples $\{P^{(sim)}_{E}(p_i)\}_i$. We point out that, despite Eq. \eqref{eq: Var PE}, the expectation value of the right-hand side of \eqref{eq: biased estimator} coincides with the left-hand side only in the limit $A \to \infty$, making $P_{E}^{(sim)}(1-P^{(sim)}_{E})$ a biased but consistent estimator for $P_{E}(1-P_{E})$. 

In order to explicitly obtain the bound $n_{min}^{(\epsilon)}(p)$ for different kinds of quantum particles, we considered the event ``both walkers in the origin'' for a 7-step quantum walk in a regime of statical percolation. Indeed, having an almost constant slope over the whole region $p\in[0,1]$, $C(p)$ is the best quantity among those studied in subsection B to estimate the value of $p$ in that regime. Notice that here the number of steps $N$ considered is reduced (with respect to the usual 15) to allow us to take a much larger value of $A$: for this analysis we randomly generated $\sim 10^5$ percolation lattices.  This guarantees that the total number of simulated configurations $A$ is much larger than the total number of possible lattices, given by $2^{7 \cdot 2} \simeq 1.6 \cdot 10^4$ for the situation considered. We stress once again that in order to obtain plots qualitatively similar to Fig. \ref{fig: cent stat}, leading to a good estimation strategy, it is important to consider an odd number of steps.

\begin{figure}
	\centering
	\subfloat[]{\includegraphics[scale=0.82]{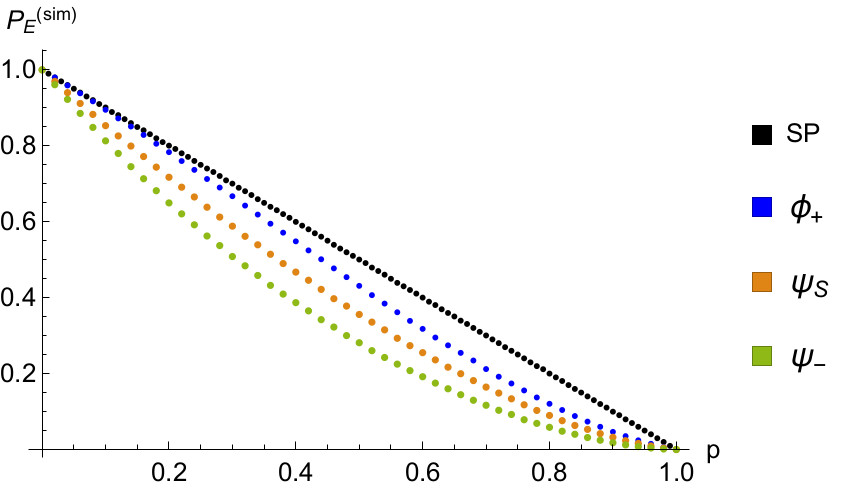}}\hskip0.5cm
	\subfloat[]{\includegraphics[scale=0.6]{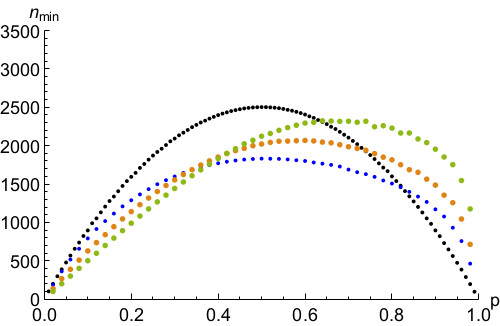}}
	\caption{$(a)$ Simulated event probability $P_{E}^{(sim)}$ of finding the single particle used (SP) or both particles [see Eq. \eqref{def: entangled states}] in the origin, as a function of the percolation parameter $p$. Due to the large number of simulations, the standard error is negligible. $(b)$ Corresponding lower bound on the number of required experimental runs $n_{min}^{(\epsilon = 0.01)}$ as a function of the percolation parameter $p$. In both cases a 7-step quantum walk has been considered, averaging over $A \sim 10^5$ lattice sequences. For the evaluation of $n^{(\epsilon)}_{min}$, we used polynomials of degree 5 to fit the points in $(a)$ (see the discussion below Eq. \eqref{eq: fit comment} in the main text). \label{fig: p estimation}}
\end{figure}

The single-walker counterpart of the event here considered corresponds to detecting the particle in the origin at the end of the walk, and it can be used to infer $p$ with the same procedure previously outlined. Therefore, we can compare the performances achieved using either one or two walkers initialised with different input exchange symmetries. In Fig. \ref{fig: p estimation} we plotted the event probability $P^{(sim)}_{E}$ and the bound $n_{min}^{(\epsilon = 0.01)}$ for the input coin states defined in Eq. \eqref{def: entangled states}, corresponding to boson-like, fermion-like statistics or to two classically indistinguishable particles. We can see how for the 7-step walk considered, for values of the percolation parameter $p\lesssim 0.82$, there is an advantage in using two quantum particles instead of one. In particular bosons are optimal in the range $0.38 \lesssim p \lesssim 0.82$, while fermions perform better below $p\simeq 0.38$. We studied the dependence of such thresholds of bosons optimality $p_1 \leq p \leq p_2$ upon $N$, considering walks with an odd number of steps between $N = 3$ and $N = 11$. We found that $p_1$ ($p_2$) decreases (increases) slightly with increasing $N$, with $p_1$ and $p_2$ varying respectively in the ranges $[0.35,0.44]$ and $[0.78,0.83]$. This suggests that the range of bosons optimality could increase with the number of steps considered.	

Finally, let us comment on the limit $n_{min}^{(\epsilon)} \to 0$ that Eq. \eqref{eq: Nmin formula} yields for $P_E \to 0$ or $P_E \to 1$.
The error $\text{VAR}\left[P_E^{(est)}\right]^{1/2}$ (and therefore $\delta p$) is associated with a certain probability of finding the real value in that range, that for a small number of measurements $n$ is actually unknown. Only for $n \gg 1$ the central limit theorem applies, allowing us to interpret the obtained error as based on a well known Gaussian probability distribution. Moreover, the linear propagation of errors assumes $\text{VAR}\left[P_E^{(est)}\right]^{1/2}$ small enough for the function $P_E(p)$ to be considered approximately linear in that interval. Even if Eq. \eqref{eq: Nmin formula} seems to suggest otherwise, the scaling $\sim 1/\sqrt{n}$ in Eq. \eqref{eq: Var PE} therefore requires to take $n_{min}^{(\epsilon)}$ large enough to satisfy the previous conditions.

\vskip0.5cm
\noindent
{\large {\bf Discussion}}

\noindent
We studied the behaviour of a one-dimensional quantum walk when the possibility of having missing bonds is considered together with the presence of a second walker. We focused on two percolation regimes, statical and dynamical, associated respectively with slowly and fast varying imperfections. Our numerical simulations have been averaged over many (typically $\sim 5000$ for a $15$-step quantum walk) different lattice sequences, realised by associating to each bond a probability $p\leq 1$ of being present. Starting from two symmetrical-spreading single-walker states, we combined them in three different ways, in order to observe the behaviour of two bosons, fermions (showing inter-particle interference), or simply two classically indistinguishable particles (with a non-symmetrised quantum state).

We described how the bosonic (fermionic) quantum statistics of the walkers affects the final probability distribution, inducing the presence of characteristic peaks (valleys) in its diagonal values. This feature, easily observed through a comparison with the output of an unsymmetrised state, is not due to percolation but emerges as a consequence of the imposed exchange symmetry between two orthogonal single-particle states. On the contrary, the remaining peaks that may still be observed in the distribution of two classically indistinguishable particles are due to a statistical effect of the average over different unitary evolutions, associated with diverse realisations of the underlying lattice structure. This shows how in the presence of percolation high probabilities of finding both walkers at the same position do not always imply a bunching phenomenon due to the bosonic nature of the particles.

The advantage of studying low-dimensional lattices with no more than two walkers is the possibility of representing the output detection probability in a 3D plot. This visual approach, however, cannot be extended to more complicated scenarios, involving multiple walkers and/or more complex structures. The search for quantities that could meaningfully describe some aspects of the walk behaviour, being at the same time easily accessible theoretically and experimentally, will therefore be of extreme importance in the future. Here we considered three of these quantities (final average distance, probability of finding both walkers at the same location or exactly in the origin) describing their qualitative behaviour for a symmetrical-spreading two-particle quantum walk with percolation. In particular we saw how they often depend strongly on the quantum statistics of the walkers even for low values of the percolation parameter $p$, fact that suggests the resilience of inter-particle interference to very noisy conditions.

Searching for multi-walkers advantages, we considered at first the spread of the walk, focusing on the dependence upon the second particle more than on the decoherence introduced by the percolation. Our analysis shows how the addition of a second walker does not increase the spread, since there always exists a single-particle state with a spread at least as large as the one obtained with two walkers. Moreover, for many cases of interest, i.e. for any couple of fermions and for bosons or classically indistinguishable particles obtained by combining two orthogonal single-particle coin states, the spread is not affected by the quantum statistics of the walkers. In these cases it is equal to the spread of the two symmetric single-particle coin states $\ket{\varphi_\pm}$ [defined in Eq. \eqref{def: symm single particle states}].

We also proposed a procedure to estimate the value of the percolation parameter that characterises the amount of lattice imperfections. Due to the monotonicity and the regularity observed in Fig. \ref{fig: cent stat}, it is worth to apply such strategy considering the detection of both particles in the origin in a regime of slowly varying imperfections, i.e., of statical percolation. For values of $p$ not too large, we showed how the performance of the proposed estimation is enhanced by the presence of a second walker, that allows us to need a smaller number of experimental measurements in order to achieve a certain estimation precision.

This investigation considered for the first time the effects of possible missing links (that could happen, for instance, in an experimental setting) on the features of walks where the quantum nature of the involved particles is relevant. Due to the importance of quantum walks for the purposes of quantum computation and simulation, and in particular for the role that many-walkers complexity will have in the future, these results pave the way for a more complete understanding and a full exploitation of these models.

\vskip0.5cm
\noindent
{\large {\bf Methods}}

\noindent
{\bf Comparison between single-particle and two-particle spread.}
Here we will provide a proof for Eq.s \eqref{prop2 statement 1} and \eqref{prop2 result}. In the following the term single-particle spread will always correspond to the quantity $V_1$ obtained from Eq. \eqref{def: variance 1 walker} by setting $i_0 = 0$ and we will use the shorthand notation $\mathcal P_i$ and $\mathcal P_{ij}$ to represent respectively the single-particle and the two-particle output probabilities to detect the walkers in positions $i$ and $j$. We will also drop the dependence upon the lattice realisation ``$a$''.

Let us begin by showing that given two orthogonal coin states $\ket{\tilde c_1}, \ket{\tilde c_2} \in H_C$, characterising the initial conditions of two single walkers localised in the origin, their average spread depends only upon the unitary evolution $U^{(N)}$ and not on the basis $\{\ket{\tilde c_1}, \ket{\tilde c_2}\}$ of $\mathcal H_C$.
Indeed, being 
	\begin{equation}
	\ketbras{\tilde c_1}{\tilde c_1}{C} + \ketbras{\tilde c_2}{\tilde c_2}{C} = \Id_C,
	\end{equation}
	from Eq. \eqref{eq: output single walker} the sum $P_i[\ket{\tilde c_1}] + P_i[\ket{\tilde c_2}]$ does not depend on the choice of basis:
\begin{equation}
P_i[\ket{\tilde c_1}] + P_i[\ket{\tilde c_2}]= \Tr\left[ U^{(N)\dagger}\ketbras{i}{i}{P}\otimes\Id_C \; U^{(N)} \ketbras{i_0}{i_0}{P}\otimes\Id_C\right],
\end{equation}
property that holds also for the average between the two single-particle spreads. Therefore,  given two particles starting from the origin with two orthonormal coin states $\ket{\tilde c_1},\ket{\tilde c_1} \in \mathcal H_C$:
\begin{equation}\label{lemma consequence}
\frac{V_1(\tilde c_1) + V_1(\tilde c_2)}{2} = V_1(\varphi_\pm),
\end{equation}
being $\braket{\varphi_+}{\varphi_-} = 0$ and $V_1(\varphi_+)=V_1(\varphi_-)$.

We are now able to prove Eq. \eqref{prop2 statement 1}.
	Being $\mathcal P_{ij}$ symmetric under the exchange $i \leftrightarrow j$, the two-particle spread $V_2$ defined in Eq. \eqref{def: variance 2W sym} only depends upon the marginal sum $\sum_j \mathcal P_{ij}$:
	\begin{equation}\label{app: spread marginal sum}
	V^{(\pm)}_2 = \sum_i i^2 \sum_j\mathcal P^{(\pm)}_{ij}.
	\end{equation}
	Using the symmetry under particle exchange of the evolved state
		\begin{equation}
		\ket{\psi^{(\pm)}_f} = \mathcal U^{(N)}_{12} \ket{\psi^{(\pm)}_{in}},
		\end{equation}
from Eq. \eqref{eq: output two walker} it follows
	\begin{align}
	\sum_j \mathcal{P}^{(\pm)}_{ij}& = \frac{1}{2}\sum_j \Tr_{1,2}\left[\left(\ketbra{ij}{ij} + \ketbra{ji}{ji}\right)\ketbra{\psi^{(\pm)}_{f}}{\psi^{(\pm)}_{f}}\right] =\Tr_1\left\{  \ketbras{i}{i}{1}\cdot U^{(N)}_1 \Tr_2\left[\ketbra{\psi^{(\pm)}_{in}}{\psi^{(\pm)}_{in}}\right]  U^{(N)\dagger}_1 \right\} \label{first line}\\
	& = \sum_{k=1}^2 \lambda_k^{(\pm)} \Tr_1\left[\ketbras{i}{i}{1}\cdot  U_1^{(N)} \ketbras{0}{0}{1}\otimes\ketbra{\tilde c_k^{(\pm)}}{\tilde c_k^{(\pm)}}  U_1^{(N)\dagger}\right] = \sum_{k=1}^2 \lambda_k^{(\pm)} \mathcal P_i\left[\ket{\tilde c_k^{(\pm)}}\right],
	\end{align}
	where the last two equalities follow from the definition of $\lambda_k^{(\pm)}$ and $\ket{\tilde c_k^{(\pm)}}$ as eigenvalues and eigenvectors of \eqref{def: lambda and c tilde}, using
	\begin{equation}\label{eqapp: form ctilde}
	 Tr_2\left[\ketbra{\psi^{(\pm)}_{in}}{\psi^{(\pm)}_{in}}\right] = \ketbras{0}{0}{P_1}\otimes \sum_k \lambda_k^{(\pm)} \ketbras{\tilde c_k^{(\pm)}}{\tilde c_k^{(\pm)}}{C_1}.
	\end{equation} 
	The required equality in Eq. \eqref{prop2 statement 1} then follows trivially from the spread definition \eqref{app: spread marginal sum}. 
	
	Eventually, it can be easily proved that for every $\ket{c_1},\ket{c_2}\in\mathcal H_C$ one has $\lambda_k^{(-)} \equiv \frac{1}{2}$, $k=1,2$, and that the same equality holds also for bosons if $\braket{c_1}{c_2} = 0$. Therefore, Eq. \eqref{prop2 result} is a consequence of Eq.s \eqref{lemma consequence} and \eqref{eq: average spread}.	

\vskip0.5cm
\noindent
{\large {\bf Acknowledgements}}

\noindent
We thank M. S. Kim for useful comments and discussions. L. R. was supported by the People Programme (Marie Curie Actions) of the European Union's Seventh Framework Programme (FP7/2007-2013) under REA grant agreement n$\degree$ 317232. C. D. F. was supported by the UK EPSRC (EP/K034480/1).

\vskip0.5cm
\noindent
{\large {\bf Author Contributions}}\\
\noindent
Both the authors made significant contributions to the concept, execution, interpretation, and preparation of the work.

\vskip0.5cm
\noindent
{\large {\bf Additional Information}}\\
\noindent
{\bf Competing Financial Interests}: The Authors declare no competing financial interests.

\vskip0.5cm
\noindent
{{\bf Corresponding authors}}: Correspondence and request of material should be addressed to Luca Rigovacca (l.rigovacca14@imperial.ac.uk) or Carlo Di Franco (cfifranc@imperial.ac.uk).

\end{document}